\documentclass[12pt,preprint]{aastex}
\slugcomment{Accepted for publication in The Astronomical Journal}
\usepackage{mathtools}
\begin{document}

\def\ntts{NTTS~155808$-$2219}
\def\kms{\ifmmode{\rm km\thinspace s^{-1}}\else km\thinspace s$^{-1}$\fi}

\title{Dynamical Measurements of the Young Upper Scorpius Triple NTTS 155808$-$2219}

\author{G. N. Mace\altaffilmark{1}, L. Prato\altaffilmark{2}, G. Torres\altaffilmark{3}, L. H. Wasserman\altaffilmark{2}, R. D. Mathieu\altaffilmark{4} and I. S. McLean\altaffilmark{1}}

\altaffiltext{1}{UCLA Division of Astronomy and Astrophysics, Los Angeles, CA 90095-1562 ; gmace@astro.ucla.edu}

\altaffiltext{2}{Lowell Observatory, 1400 West Mars Hill Road, Flagstaff, AZ 86001}

\altaffiltext{3}{Harvard-Smithsonian Center for Astrophysics, 60 Garden St., Cambridge,
MA 02138}

\altaffiltext{4}{Department of Astronomy, University of Wisconsin - Madison,
Madison WI 53726}

\begin{abstract}

The young, low-mass, triple system \ntts\ (ScoPMS 20) was previously identified as a $\sim$17-day period single-lined spectroscopic binary with a tertiary component at 0.21 arcseconds. Using high-resolution infrared spectra, acquired with NIRSPEC on Keck II, both with and without adaptive optics, we measured radial velocities of all three components. Reanalysis of the single-lined visible light observations, made from 1987 to 1993, also yielded radial velocity detections of the three stars. Combining visible light and infrared data to compute the orbital solution produces orbital parameters consistent with the single-lined solution and a mass ratio of q = 0.78 $\pm$ 0.01 for the SB. We discuss the consistency between our results and previously published data on this system, our radial-velocity analysis with both observed and synthetic templates, and the possibility that this system is eclipsing, providing a potential method for the determination of the stars' absolute masses. Over the $\sim$20 year baseline of our observations, we have measured the acceleration of the SB's center-of-mass in its orbit with the tertiary. Long-term, adaptive optics imaging of the tertiary will eventually yield dynamical data useful for component mass estimates.

\end{abstract}


\section{Introduction}
To understand the star formation process, reliable observations of fundamental stellar properties are needed so theoretical models can be tested. From one star forming region (SFR) to the next the binary fraction can vary \citep{2000prpl.conf..703M} and may depend on the density of the region \citep{1997ApJ...482L..81S,2003ApJ...583..358B} and spectral type of the primary \citep{2006ApJ...640L..63L}, underscoring the importance of a complete binary census. Characterization of multiples' frequency, component separation distribution, and mass ratio distribution for a given SFR provides clues to the broad star forming properties (angular momentum, density, turbulence, etc.) of the parent molecular cloud. Small separation spectroscopic binaries with periods sufficiently short to enable the measurement of the individual stellar velocities, and thus of the system's mass ratio, are potential targets for the dynamical determination of individual component stellar masses \citep[e.g.,][]{2001AJ....122..997S,pra02a,2005ApJ...635..442B}. Knowledge of absolute masses and observable properties, such as effective temperature and luminosity, plays a key role in the improvement of pre-main sequence evolutionary models \citep[e.g.,][]{bar98,2001ApJ...553..299P}. This work is part of our effort to measure a substantive sample of young star mass ratios and masses with the ultimate goals of improving our understanding of tight binary formation and of tying models of young star evolution to concrete dynamical data.  \ntts\ is a particularly important target given its high-order multiplicity and its low-mass components, a regime plagued by discrepancies in theoretical calculations.

Spectroscopic binaries (SBs) with low mass-ratios are usually identified as single-lined spectroscopic systems when observed in visible light because the large difference in flux between the primary and secondary at short wavelengths prevents detection of the secondary component (Mazeh et al. 2002). In the Raleigh-Jeans regime of the stars' spectral energy distribution, however, flux scales much less steeply as a function of mass. Thus, by observing single-lined spectroscopic binaries with infrared (IR) spectroscopy we are able to improve our chances of detecting the lower-mass secondary not seen in visible light, as initially described in \citet{1998PhDT........16P}, \citet{pra02b}, and \citet{maz02,2003ApJ...599.1344M}.

Observations of {\it Einstein} x-ray sources in Upper Scorpius by \citet[ hereafter W94]{1994AJ....107..692W} identified the spectral type M3 \ntts\ system as an SB with a tertiary component (SB$+$T). Figure 11 of W94 presents a visible light spectrum with three distinct Li $\lambda6707$ \AA\ lines. Based on this evidence for multiple components, they estimated an SB mass ratio of 0.71 $\pm$ 0.05. \citet{1994ARA&A..32..465M} provided initial single-lined orbital parameters for the \ntts SB, revealing a 16.925 day period, -5.0~\kms\ center-of-mass velocity ($\gamma$), consistent with membership in Upper Sco, and an eccentricity of 0.10. \citet{2000A&A...356..541K} later determined an SB$+$T separation of 0.193$''$, which is consistent with the value of 0.21$''$ determined by \citet{2001A&A...369..249W}. Equivalent widths of the 6707\AA~Li line ($\sim$0.6\AA) and H$\alpha$ ($\sim$ -3\AA) suggest a young age and a low accretion rate for the system.  The lack of evidence for circumstellar dust, based on 3.4-22 $\mu$m photometry from the Wide-field Infrared Survey Explorer (WISE), is consistent with a  weak-lined T Tauri star (WTTS) classification. Observed properties of \ntts\ are presented in Table~\ref{tbl-1}.

This paper describes the results of combining high-resolution IR spectroscopy with reprocessed visible light data to determine the component radial velocities of \ntts\ over a nearly 20 year timespan. In \S2 we briefly describe our observations and data reduction. Our cross-correlation analysis appears in \S3, and the \ntts\ orbital solution is presented in \S4. Section 5 contains our discussion of the system age and component masses, the possibility of an eclipse, given the large SB orbital inclination, and initial estimates for the tertiary orbital characteristics. A summary of our findings appears in \S6.

\section{Observations and Data Reduction}

\subsection{Visible Light Spectroscopy}
Spectroscopic observations of \ntts\ were collected from 1987 June to 1993 April on the 37 epochs listed in Table~\ref{tbl-2} (W94). Two nearly identical echelle spectrographs, with photon-counting intensified Reticon detectors, were used on the Multiple Mirror Telescope on Mount Hopkins, Arizona (prior to its conversion to a monolithic mirror), and on the 1.5-m Tillinghast reflector at the F.\ L.\ Whipple Observatory, also atop Mount Hopkins. A single echelle order, 45\ \AA\ wide, was recorded at a central wavelength of 5188.5\ \AA, containing the \ion{Mg}{1}~b triplet. The resolving power provided by this setup was R$=\lambda/\Delta\lambda\approx 35,\!000$. The nominal signal-to-noise ratios of the 37 spectra obtained range from $\sim$6 to 19 per resolution element of 8.5~\kms. Standard flatfield frames were obtained each night, and the pixel-to-wavelength mapping was based on exposures of a thorium-argon lamp taken before and after each science exposure \citep[see][]{Latham:02}.

\subsection{Near-Infrared Spectroscopy}
Observations were made in the IR between 2000 June and 2007 April with the Keck~II 10-m telescope on Mauna Kea. The UT dates of observation are listed in Table~\ref{tbl-3}. Nine $H$-band orders, with a central wavelength of $\sim$1.555 $\mu$m in the middle order, were obtained using the facility near-infrared, cross-dispersed, cryogenic spectrograph NIRSPEC \citep{mcl98, mcl00}. NIRSPEC employs a 1024 $\times$ 1024 ALADDIN InSb array detector. Source acquisition was accomplished with the slit viewing camera, SCAM, which utilizes a 256 $\times$ 256 HgCdTe detector. \ntts\ has a 2MASS $H$-band magnitude of 9.01 (Table 1); integration times for individual frames were 300 seconds with the NIRSPEC-5 filter. We nodded the telescope to move the star between two positions on the slit, separated by $\sim$60~pixels, to allow for background subtraction by differencing sequential spectra.

\begin{figure*}
\includegraphics[angle=0,width=6.0in]{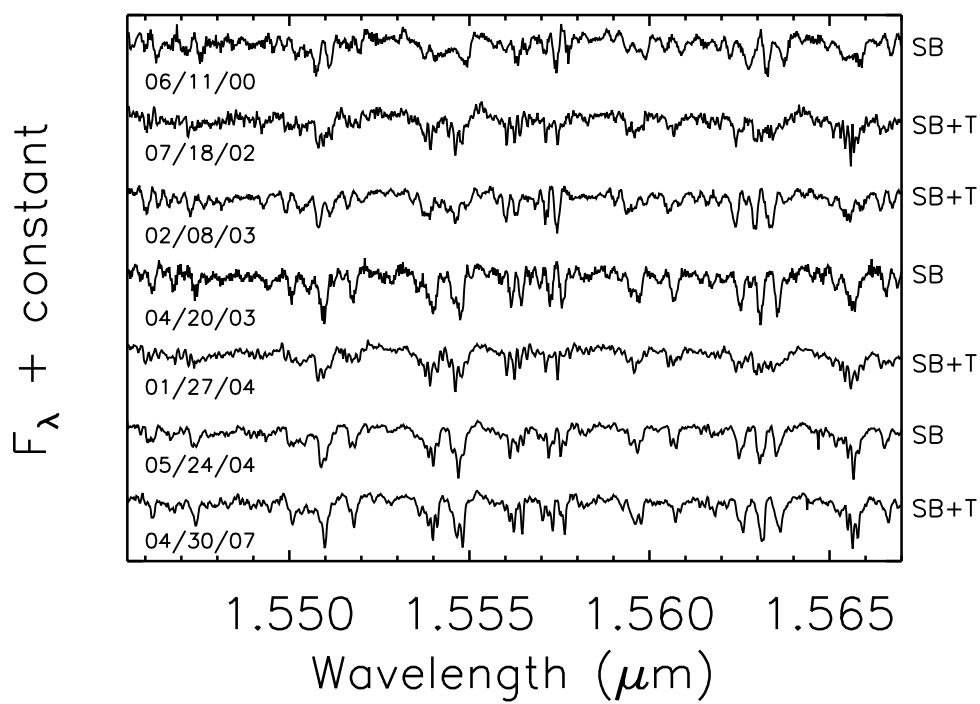}
\caption{NIRSPEC H-band spectra of \ntts. The three epochs of AO observations are SB only, while non-AO spectra include the SB+Tertiary. All spectra have been corrected for heliocentric motion, flattened, normalized to a continuum level of unity, and separated in the y-axis by an additive constant.} 
\label{fig1}
\end{figure*}

Four of the seven observations used the 24$''$ $\times$ 0.288$''$ slit and contain light from the three SB+T components. The Keck Adaptive Optics (AO) system effectively reimages the optics, reducing the plate scale by a factor of $\sim$11, producing slit dimensions of 2.3$''$ $\times$ 0.027$''$. Three IR observations utilized AO in front of NIRSPEC. On all three AO nights we obtained isolated spectra of the SB. On two of these occasions we utilized SCAM imaging to align the slit to also include the tertiary. All spectra have resolution R~$\approx 30,\!000$. The SB and SB$+$T spectra from the seven epochs of observation are shown in Figure~\ref{fig1} for a single H-band order.

All spectroscopic reductions were made by using the REDSPEC package, software written at UCLA by S. Kim, L. Prato, and I. McLean specifically for the analysis of NIRSPEC data\footnote{See: http://www2.keck.hawaii.edu/inst/nirspec/redspec/index.html}, following the procedures outlined by \citet{pra02b}. We used the central order, 49, found at our setting. This order has three advantages: (1) it is rich in both atomic and molecular lines and is therefore suitable for identifying spectra of both warm and cool stars, (2) the OH night sky emission lines across order 49 are numerous and well-distributed, yielding accurate dispersion solutions, (3) this order has the advantage of lacking prominent telluric absorption lines. Consequently, we did not have to divide our target spectra by telluric star spectra. Because the AO reimaged slit width is so narrow, it was not possible to use OH night sky emission for wavelength calibration on the AO nights.  On these occasions we took comparison lamp exposures at the beginning and/or end of the night to determine the dispersion solution and zero point.

\section{Analysis}
\subsection{Visible Light Radial Velocities}
Based on the expectation that the secondary would be too weak to detect in our spectra, and that only the primary and tertiary components would be visible, we initially derived radial velocities for only these two components of the \ntts\ system using TODCOR, a two-dimensional cross-correlation technique \citep{Zucker:94}. The same synthetic template was used for the primary and tertiary stars, based on Kurucz model atmospheres \citep{Latham:02}, with a temperature of 3750~K selected in accordance with the late spectral type of the unresolved system, M3, and zero rotational broadening, vsin$i$ = 0 \kms. The large scatter in the tertiary velocities suggested possible contamination either from moonlight in a few exposures, or more likely by the secondary component. A reanalysis using an extension of TODCOR to three dimensions \citep{Zucker:95} showed that the lines of the secondary are indeed visible in most of our visible light spectra, and radial velocities were derived using a template for the secondary identical to that of the primary and tertiary. The moonlight contamination was removed from seven of the exposures using a four-dimensional version of TODCOR \citep{Torres:07} and the addition of a fourth template appropriate for the Sun. The radial velocities are listed in Table~\ref{tbl-2}, and have typical uncertainties of 1.7~\kms\ for the primary, 4.4~\kms\ for the secondary, and about 3.5~\kms\ for the tertiary. The stability of the zero-point of the CfA velocity system was monitored by means of exposures of the dusk and dawn sky, and small systematic run-to-run corrections were applied in the manner described by \citet{Latham:92}, already included in the velocities described above. In the case that either the secondary or the tertiary radial velocity was impossible to determine, only the primary star velocity was used. This resulted in 37 primary, 33 secondary, and 33 tertiary velocities. The light ratios among the three stars determined from our spectra are $\alpha~=\ell_2/\ell_1 = 0.30 \pm 0.03$ and $\beta~= \ell_3/\ell_1 = 0.53 \pm 0.04$ at the mean wavelength of our observations. The tertiary companion is thus brighter than the SB secondary. Comparison of the visible light spectra to GL 752A, a main sequence star with vsin$i$~$\le$~2.5~\kms\ \citep{2010AJ....139..504B}, implies a low, $\le$~10~\kms, rotational velocity for all components of \ntts.

\subsection{Infrared Radial Velocities}

Stellar radial velocities were also measured for the IR spectra using a similar multi-dimensional cross-correlation analysis following the approach of \citet{Zucker:94}. To optimize the correlation, and thus obtain the most precise radial velocities, the best matching synthetic or observed templates must be used for cross-correlation with the target observation.  In our analysis of the IR data we tested three sets of templates:  (1) the \ntts\ tertiary spectrum, (2) synthetic spectra, and (3) observed spectra of radial velocity standards.

The tertiary spectra isolated in the AO observations on UT 2000 Jun 11 and 2003 Apr 20 provided an excellent template because this bound companion likely shares the same age, metallicity, and surface gravity with the close SB components. Assuming a distance of 145~pc to the region, the $\sim$0$\farcs$21 angular separation of the tertiary implies a projected physical separation of $\sim$30~AU. This ability to angularly resolve the subarcsecond, visual companion illustrates a unique advantage of IR observations. Cross-correlation of the AO tertiary spectra against one another revealed no significant change in radial velocity, within the $\sim$1\kms\ uncertainty \citep{pra02b,2001AJ....122..997S}. Thus the tertiary spectra were combined to use as a cross-correlation template with the highest possible SNR.

\begin{figure*}
\includegraphics[angle=0,width=6.0in]{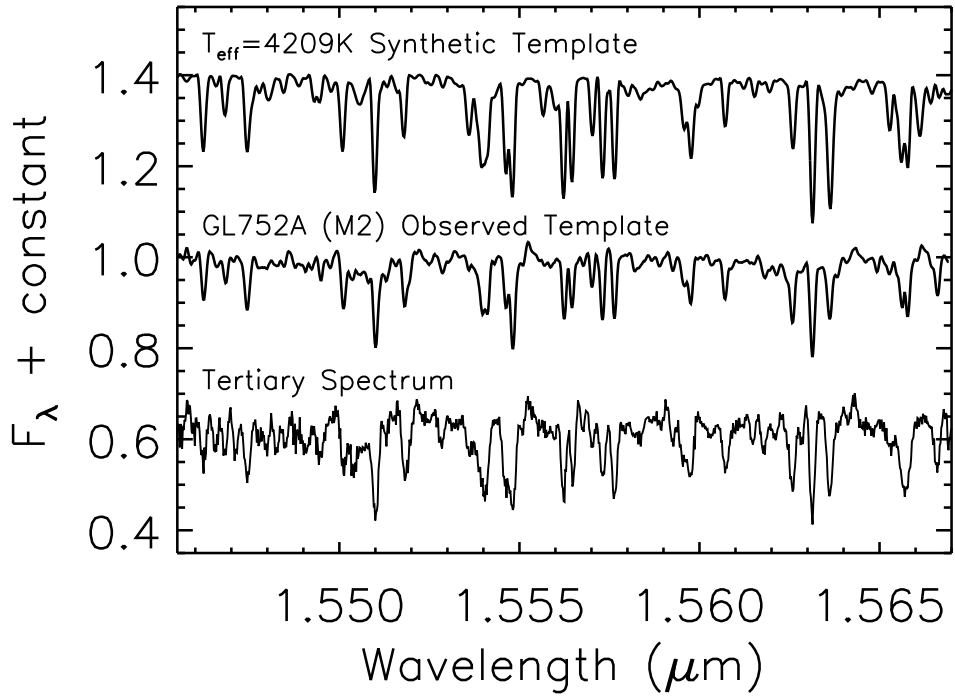}
\caption{Spectra of the \ntts\ tertiary and cross-correlation templates. Top: Best matching synthetic spectrum calculated with the PHOENIX model atmosphere code and rotationally broadened to 25~\kms. Middle: Best matching observed template, GL 752A, rotationally broadened to 18~\kms. Bottom: Averaged spectrum from the two NIRSPEC+AO epochs (UT 2000 June 11 and 2003 April 20) which isolated the tertiary. All spectra have been corrected for heliocentric motion,  flattened, normalized to a continuum level of unity, and separated in the y-axis by an additive constant. Our analysis shows that the GL752A spectrum provides the highest correlation, while the synthetic template gives more consistent radial velocities and correlation values within 5$\%$ of the GL 752A results.} 
\label{fig2}
\end{figure*}

The observed template that best matches the \ntts\ tertiary IR spectrum is the main-sequence star GL 752A. In a comparison of a sequence of high-resolution, near-IR spectra, \citet{2007ApJ...657..338P} found a spectral type of M2 for GL 752A, similar to the M2.5 type of \citet{2010AJ....139..504B} based on visible light observations; SIMBAD lists a type of M3. We adopt M2 throughout our analysis. Figure~\ref{fig2} shows the GL 752A spectrum together with the best matching synthetic template and the average of the two \ntts\ tertiary spectra taken on AO nights. 

Our synthetic spectral library was generated from atmospheric structures calculated with the PHOENIX model atmosphere code \citep{1999ApJ...512..377H} and kindly provided by T. Barman. Synthetic templates were compared to observed templates by cross-correlation. \citet{2007ApJ...657..338P} determined the spectral types of a subset of our observed templates; the corresponding effective temperatures (T$_{eff}$) were estimated from \citet{2003ApJ...593.1093L}.  For the observed templates, the highest correlations were obtained with synthetic templates of typically $\sim$500 K hotter T$_{eff}$, a known characteristic of model atmospheres \citep{2005MNRAS.358..105J}.  Because we derive the radial velocity from the cross-correlation, this T$_{eff}$ discrepancy doesn't impact the radial velocities measured with the synthetic templates.  Four synthetic templates were ultimately used, T$_{eff}=$ 3645, 3883, 4209, and 4831 K, corresponding to 1 Gyr old models of 0.5, 0.6, 0.7, and 0.8 M$_{\odot}$ stars with log$g$ of $\sim$4.7 \citep{bar98}. The T$_{eff}=4209$ K synthetic spectrum yielded the highest correlation with the \ntts\  tertiary.

The GL 752A spectrum was convolved with a series of kernels to mimic a range of rotational velocities, as described in \citet[][section 4.1]{2007ApJ...657..338P}.  The highest correlation with the \ntts\ tertiary was obtained for $v$sin$i=22 \pm1$ km/s.  However, radial velocities derived with the GL 752A template rotated to other values of $v$sin$i$ vary by a few km/s.  Cross-correlation of the T$_{eff}=4209$ K synthetic template against the tertiary resulted in correlation coefficients that are $\sim$5\% lower than for the observed GL 752A template, but yielded radial velocities consistent over a wide range of rotational velocities.  The highest correlation was obtained with the T$_{eff}=4209$ K spectrum rotated to 24 km/s.

All three types of templates were used to analyze the 2003 April 20 and 2004 May 24 SB spectra, for which the component radial velocity separation was maximum and the SB isolated. Using the averaged tertiary spectrum as a template for both the primary and secondary resulted in the highest correlation. GL 752A as a template for the primary and secondary produced nearly the same correlation as the tertiary as the template. However, for a range in vsin$i$'s, the most consistent velocities were again found using the synthetic templates. When the various templates were applied to the SB+T spectra, only the synthetic templates produce reliable radial velocities, similar flux ratios to the SB-only analysis, and high correlations.

Testing different synthetic templates to find the maximum correlation confirmed that the T$_{eff}$ = 4209~K spectrum produced the best match to the primary and tertiary, while the secondary was best fit by the T$_{eff}$ = 3883~K synthetic spectrum. $V$sin$i$ values of 20 $\pm$ 1~\kms, 24 $\pm$ 4~\kms, and 24 $\pm$ 3~\kms\ for the primary, secondary, and tertiary, respectively, yielded the highest correlations. Vsin$i$ uncertainties are the standard deviation of the highest correlated templates for all IR epochs. Our visible light cross-correlation analysis yielded vsin$i$ values of $\le$~10~\kms. Because the visible light synthetic templates used in this analysis were created with inherent rotation, rather than through convolution with a kernel function after the spectra were synthesized, we suspect that the lower vsin$i$ values are likely more accurate. We therefore adopt vsin$i$=10 km/s for all three components of \ntts.

The radial velocities determined with the synthetic template spectra are listed in Table~\ref{tbl-3}, and have typical uncertainties of 1.9~\kms\ for the primary, 2.6~\kms\ for the secondary, and 1.8~\kms\ for the tertiary.  The uncertainty in each radial velocity was determined as a function of the FWHM of the correlation peak, and the ratio of the correlation peak height to the amplitude of the antisymmetric noise, following the procedure of \citet{1992ASPC...25..432K}. Ideally we would use a three dimensional cross-correlation for all SB+T spectra but the large flux ratios and similar spectral types prevented convergence on a solution. In these cases we used a two dimensional cross-correlation to fit the primary and tertiary components, then identified the secondary radial velocity as a residual peak in the cross-correlation function.

The three pure SB spectra observed with NIRSPEC+AO yielded an H-band light ratio of the secondary/primary, $\alpha = 0.55 \pm 0.10$. For the four SB+T spectra, $\alpha$ was held at 0.55 during cross-correlation. The 2002 July 18 and 2004 January 27 SB+T spectra were used to determine the tertiary/primary light ratio, $\beta = 0.81 \pm 0.09$, and for cross-correlation of the four SB+T spectra $\beta$ was held at 0.81. H-band magnitudes from \citet{2001A&A...376..982W} produce the photometric H-band brightness ratio T/SB = 0.58 $\pm$ 0.03, consistent with our T/SB H-band value of 0.52 $\pm$ 0.13.

\section{SB Orbital Parameters}  

\begin{figure*}
\includegraphics[angle=0,width=6.0in]{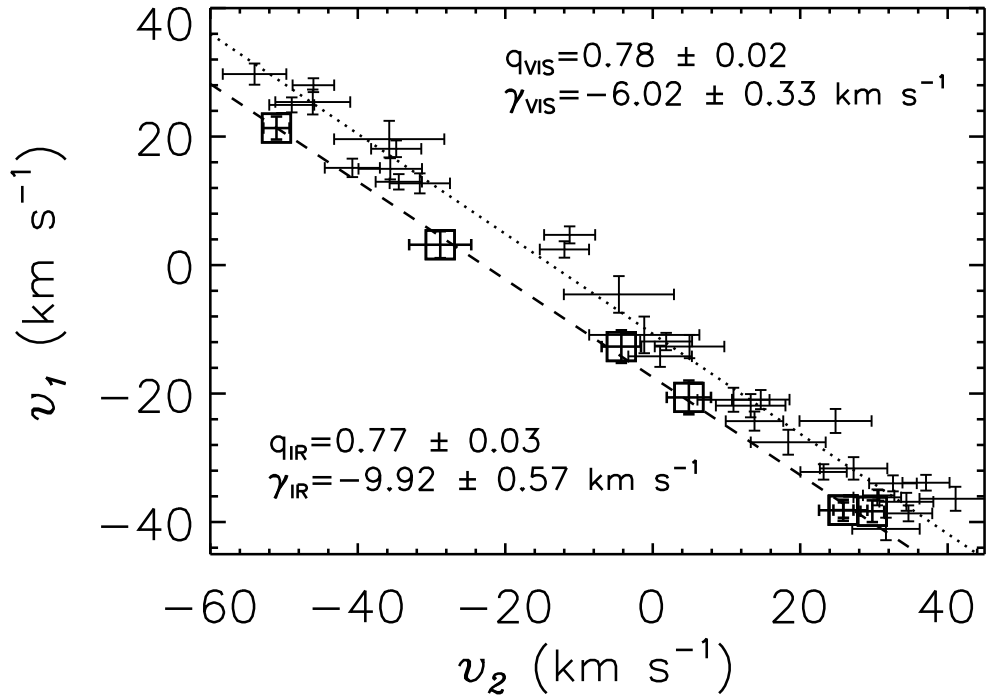}
\caption{The primary vs$.$ secondary radial velocities for \ntts\ following \citet{wil41}. The dotted line shows the weighted fit to the visible light data (bare error bars) and the dashed line shows the fit to the IR velocities (boxes). The mass ratio, q, is the negative of the slope of the fit and the center-of-mass velocity is determined by the equation $\gamma$ = (y-intercept)/(1 + q). The vertical offset of the two fits gives a center-of-mass acceleration of d$\gamma$/dt = -0.30 $\pm$ 0.07 (\kms\thinspace yr$^{-1}$).} 
\label{fig3}
\end{figure*}

Preliminary estimates for the center-of-mass velocity and mass ratio were derived using the method of \citet{wil41}. Figure~\ref{fig3} shows the primary velocity vs$.$ the secondary velocity, for the epochs from which both were determined for the double-lined spectra, along with the weighted linear fits for the visible light and IR velocities, separately. From the linear fit, the negative of the slope gives the mass ratio, q. A very similar value, to within 1$\sigma$, was obtained for the visible light and IR data sets. The center-of-mass velocity $\gamma$ is given by the y-intercept divided by (1+q). Although the full orbital solution provides more precise measurements of these quantities by adding phase constraints, the vertical offset between the two data sets produces a potentially significant measurement of the center-of-mass acceleration. For the 13.18 years between the mean epochs of observation d$\gamma$/dt = -0.30 $\pm$ 0.07 (\kms\thinspace yr$^{-1}$).

\begin{figure*}
\includegraphics[angle=0,width=6.0in]{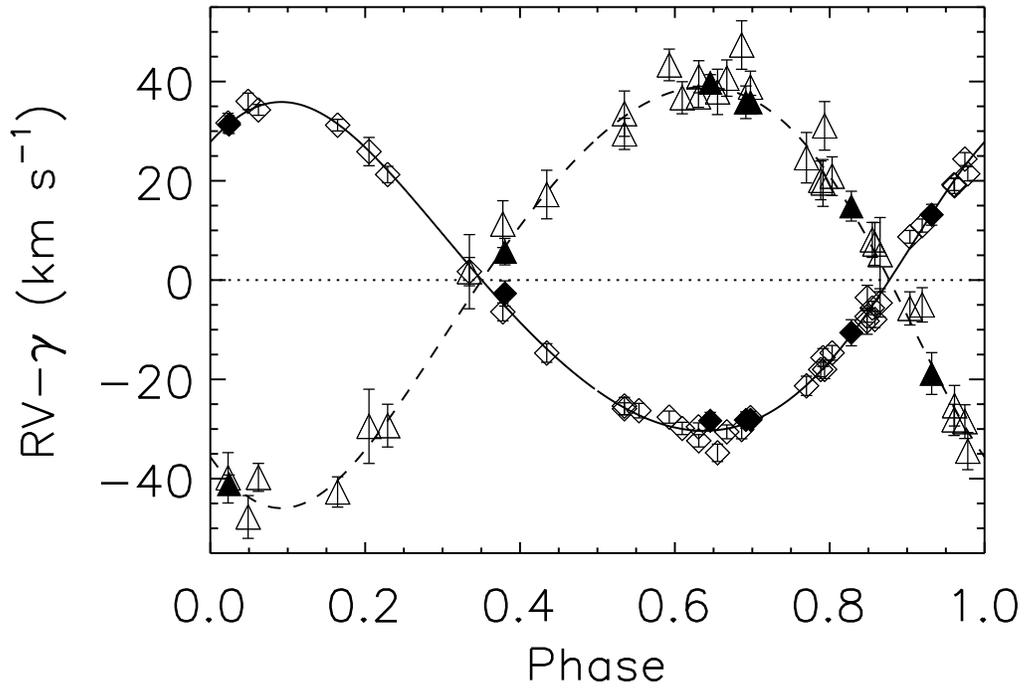}
\caption{Radial velocity minus the center-of-mass velocity as a function of phase for \ntts. The diamonds represent the primary star data, and the triangles the secondary star data. Filled symbols are IR velocities and open symbols are visible light velocities. The best fit orbital solution to the combined visible light and IR data is represented by a solid line for the primary star and a dashed line for the secondary star.} 
\label{fig4}
\end{figure*}

Table~\ref{tbl-4} lists the orbital parameters derived by \citet{1994ARA&A..32..465M} and W94, along with the solutions determined using our visible light and IR radial velocities. The orbital fit to the radial velocities of both SB components is accomplished with a standard least-squares analysis using the Levenberg-Marquardt method \citep{1992nrfa.book.....P}. Initial guesses for the solution were found by using an amoeba search routine also from \citet{1992nrfa.book.....P}. Refinements to the initial guess orbital elements were made to ensure that the best-fit, in the $\chi^2$ sense, was consistent with the single lined solution of \citet{1994ARA&A..32..465M}. We determined the period, P, the projected semi-major axes of the components, $a_1$sin$i$ and $a_2$sin$i$, eccentricity, e, periastron angle, $\omega$, the time of periastron passage, T, the center-of-mass velocity, $\gamma$, and the center-of-mass acceleration, d$\gamma$/dt. From the fit to the phased radial velocity curves (Figure~\ref{fig4}) we can measure the semi-amplitude of both components, $K_1$ and $K_2$, and derive the mass ratio, q = $K_1$/$K_2$ = $a_1$sin$i$ / $a_2$sin$i$, and the minimum masses, $M_1$sin$^3$$i$ and $M_2$sin$^3$$i$. Uncertainties calculated by the Levenberg-Marquardt method are the formal errors from the nonlinear least-squares fitting. The orbital elements derived using all the data are dominated by the more numerous visible light radial velocities and are consistent with the individual IR and visible light orbital solutions. 

\begin{figure*}
\includegraphics[angle=0,width=6.0in]{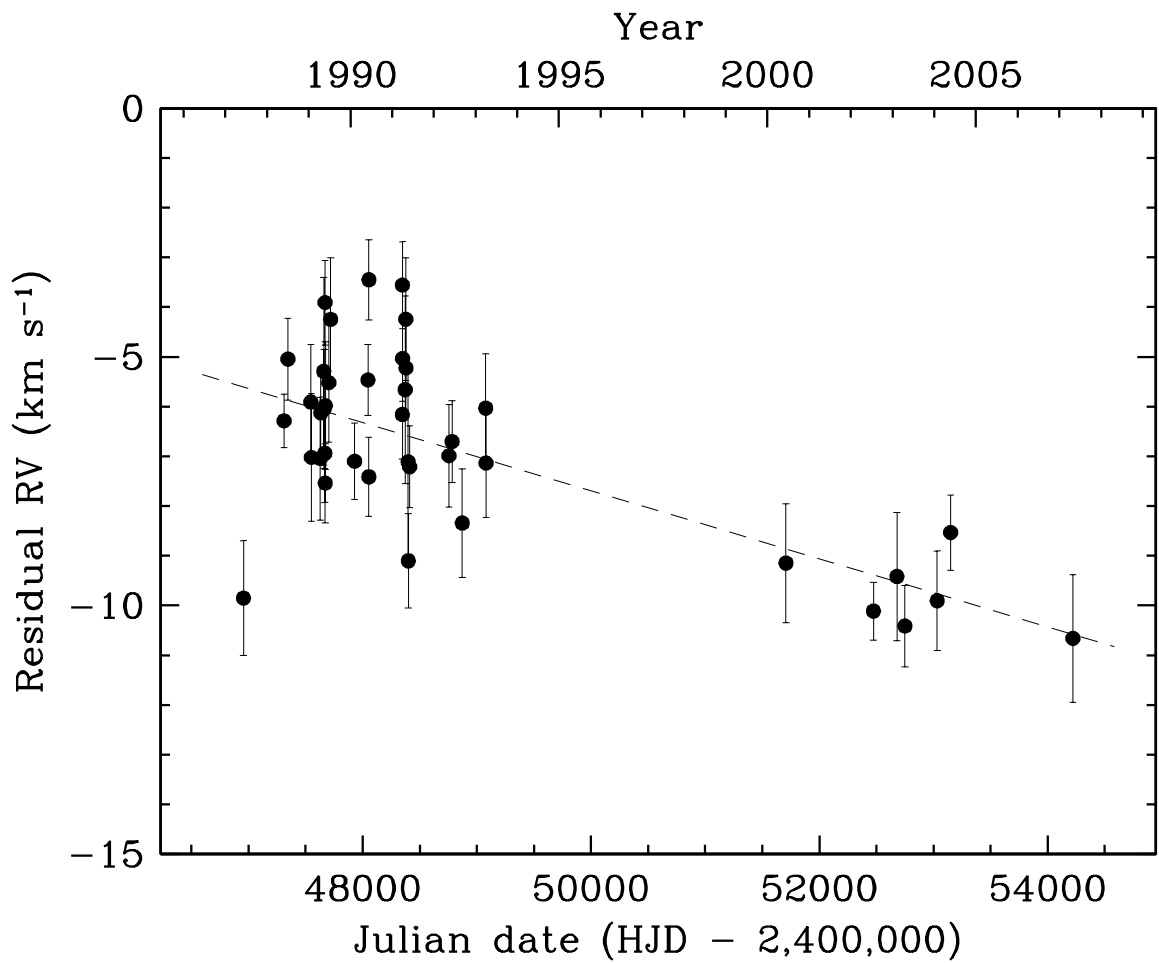}
\caption{Change in the center-of mass velocity of the binary as a function of time. Each point represents the weighted average of the primary and secondary residuals after subtracting the orbital motion. The dashed line corresponds to the fitted $\gamma$ velocity from our orbital fit as a function of time. The center-of-mass acceleration from the orbital solution is d$\gamma$/dt = -0.25 $\pm$ 0.04 (\kms\thinspace yr$^{-1}$), which is consistent with the method in Figure~\ref{fig3}.} 
\label{fig5}
\end{figure*}

On the basis of the combined visible and IR data we find that the SB in \ntts\ has an orbital period P = 16.9243 $\pm$ 0.0002 days, center-of-mass acceleration d$\gamma$/dt = -0.25 $\pm$ 0.04 (\kms\thinspace yr$^{-1}$), and eccentricity e = 0.113 $\pm$ 0.008. Our center-of-mass velocity, $\gamma$ = -8.06 $\pm$ 0.29 \kms\ is consistent with the value of $\gamma$ = -6.28 $\pm$ 3.04 \kms\ for Upper Scorpius M dwarfs surveyed by \citet{2012ApJ...745...56D}.The combined orbital solution, as a function of phase, is plotted with the measured radial velocities in Figure~\ref{fig4}. In order to depict the combined orbital solution as a function of phase, $\gamma$ has been subtracted from each of the measured radial velocities. The mass ratio, derived from the ratio of projected semi-major axes, is q = 0.78 $\pm$ 0.01, consistent with the preliminary analysis based on Figure~\ref{fig3}. The drift in the center-of-mass velocity of the binary is illustrated in Figure~\ref{fig5}, where we have subtracted the orbital motion from the individual velocities, and represented the weighted average of the residuals as a function of time. The dashed line corresponds to the change in $\gamma$ from the joint orbital fit.

\section{Discussion}

\subsection{Component Masses and System Age}
The template analysis in \S 3.2 revealed that the observed GL 752A template generally produced the highest correlation for all three components and is best matched to the primary, which we classify as an M2$\pm$1. Cross-correlation using GL 436 (M3) as the secondary template resulted in a higher correlation than GL 752A for the 2003 April 20 SB spectrum. The secondary was also fit better by a lower temperature synthetic template, resulting in our classification of the secondary as an M3$\pm$1. Although the GL 752A and T$_{eff}$ = 4209~K templates provided the highest correlation coefficients for the tertiary, the T$_{eff}$ = 3883~K and GL 436 templates were similar to within a few percent. Thus, we classify the tertiary at an intermediate spectral type of M2.5$\pm$1. Although the synthetic templates yielded more consistent RVs, because of the known discrepancy in T$_{eff}$, i.e. the T$_{eff}$ values of synthetic spectra are anomalously hot, the observed templates provided a more accurate method for estimating the \ntts\ spectral types and hence T$_{eff}$s.

To place \ntts\ on an H-R diagram, we use the spectral types M2, M3, and M2.5 (\S 3.2) and the temperature scale for young stars from \citet{2003ApJ...593.1093L}: we found T$_{eff}$ = 3560 $\pm$ 145,  3415 $\pm$ 145, and 3488 $\pm$ 145~K for the primary, secondary and tertiary, respectively. We assume a distance to Sco-Cen of 145 $\pm$ 15 pc \citep{1999AJ....117..354D}. Estimates of A$_\nu$ in the literature are 0.3 (W94) and 0.5 magnitudes \citep{1999AJ....117.2381P}. Using 2MASS magnitudes and the color-color method described in \S 3.1 of \citet{2003ApJ...584..853P} yields A$_\nu$ = 0.47. We adopt A$_\nu$ = 0.5$\pm$0.2 in our analysis. Finally, using data from 2MASS and the H-band flux ratios determined by cross-correlation (\S 3.2), we calculated the luminosity of the primary, log(L$_1$/L$_\odot$) = -0.56 $\pm$ 0.04, secondary, log(L$_2$/L$_\odot$) = -0.84 $\pm$ 0.04, and tertiary, log(L$_3$/L$_\odot$) = -0.66 $\pm$ 0.04 following the approach in \S 3.3 of \citet{2003ApJ...584..853P} and using d$=$145 pc. The \ntts\ system lacks any detectable near-IR excess (i.e. r$_k$=0; Prato et al. 2003). Table~\ref{tbl-5} lists these derived properties.

\begin{figure*}
\includegraphics[angle=0,width=6.0in]{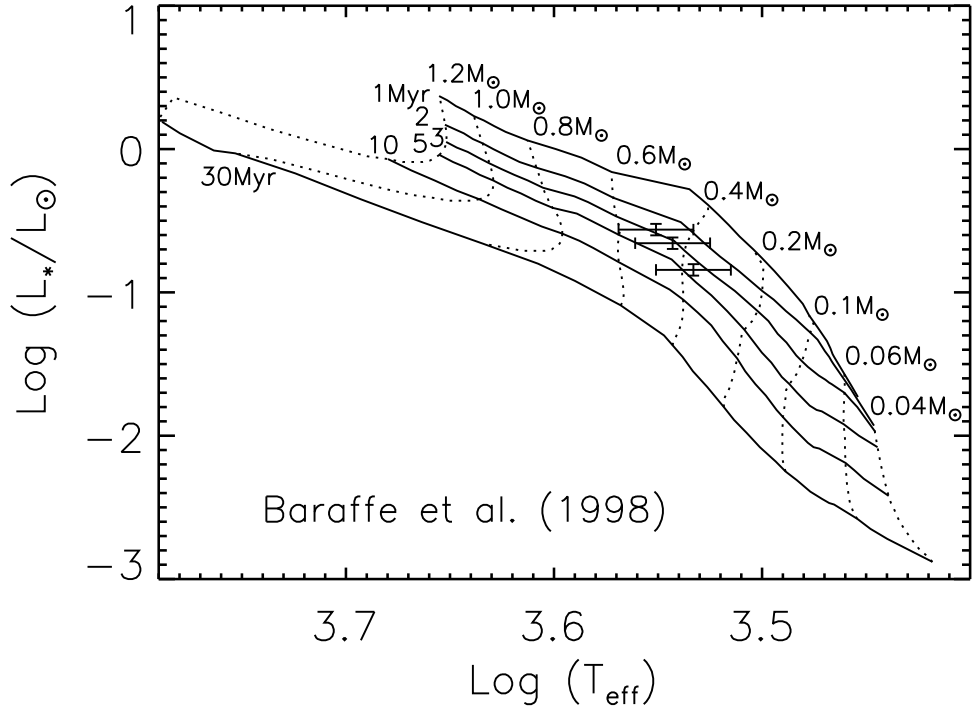}
\caption{T$_{eff}$ and luminosity for the \ntts\ components plotted on the pre-main sequence evolutionary tracks of \citet{bar98} for solar metallicity and mixing lengths $\alpha$ = 1.0 for $M < 0.5 M_{\sun}$ and $\alpha$ = 1.9 otherwise. Luminosities for the three components were derived from the total 2MASS magnitudes for the system and the spectroscopic component flux ratios. Effective temperatures were chosen from \citet{2003ApJ...593.1093L} for an M2 $\pm$ 1 primary, M3 $\pm$ 1 secondary and M2.5 $\pm$ 1 tertiary. The mass tracks and isochrones are labeled. For objects lying between labelled isochrones and mass tracks, the ages and masses were estimated by interpolation. The three \ntts\ components appear to be coeval with an age of 3.3 $^{+2.2}_{-1.3}$ Myr. } 
\label{fig6}
\end{figure*}

\begin{figure*}
\includegraphics[angle=0,width=6.0in]{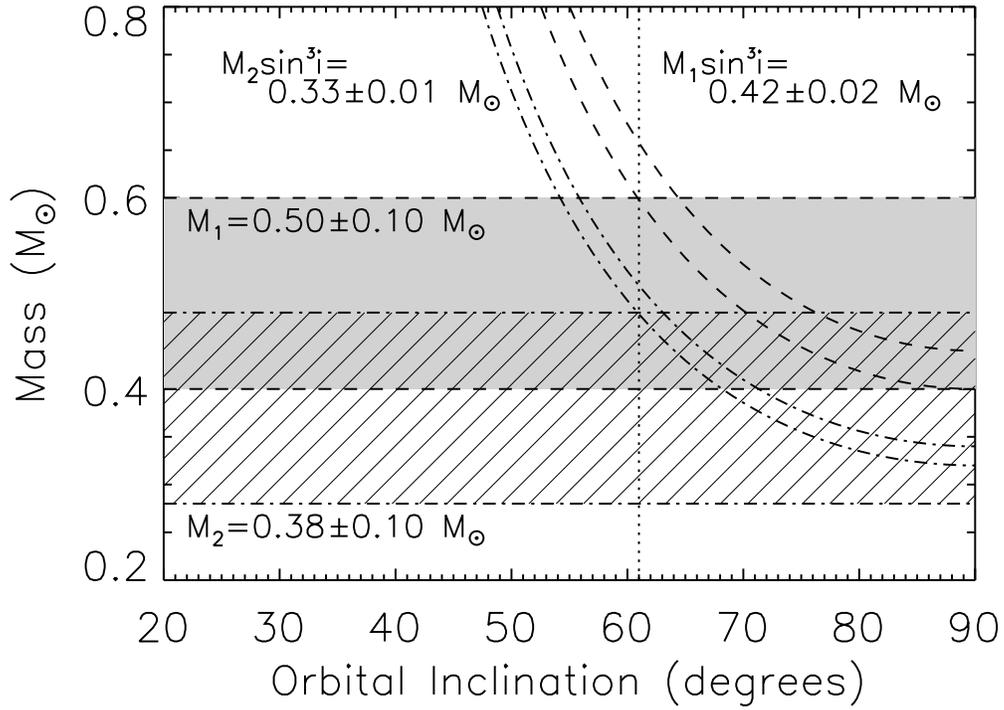}
\caption{Masses of the \ntts\ SB components as a function of orbital inclination. The values of $M_1$sin$^3 i$ and $M_2$sin$^3 i$ derived from the orbital solution are shown with 1$\sigma$ uncertainties (curved dashed and dash-dot lines). Primary and secondary mass estimates from the tracks of \citet{bar98} are over-plotted (horizontal dashed and dash-dot lines), also with 1$\sigma$ uncertainties. The inclination must be $>$ 61$^o$ for the model masses to be consistent with the minimum masses within the 1$\sigma$ uncertainties.} 
\label{fig7}
\end{figure*}

The positions of the \ntts\ components on the pre-main sequence evolutionary tracks of \citet{bar98} are shown in Figure~\ref{fig6}. From this plot we are able to estimate the masses of the primary, M$_1$ = 0.50 $\pm$ 0.10 M$_\odot$, secondary, M$_2$= 0.38 $\pm$ 0.10 M$_\odot$, and tertiary, M$_3$ = 0.44 $\pm$ 0.10 M$_\odot$. These masses give a secondary-to-primary mass ratio M$_2$/M$_1$ = 0.76 $\pm$ 0.14, consistent with our orbital solution, a tertiary-to-primary mass ratio M$_3$/M$_1$ = 0.88 $\pm$ 0.14, and a tertiary-to-SB mass ratio M$_T$/M$_{SB}$ = 0.50 $\pm$ 0.17. Using the K-band flux ratio from \citet{2000A&A...356..541K} as a proxy for the mass ratio, as in \citet{2008ApJ...679..762K}, results in a T/SB mass ratio of 0.57 $\pm$ 0.10, which is in agreement with our estimate. We combined these mass estimates and our orbital solution (Table~\ref{tbl-4}) and followed the approach of \citet{2001ApJ...549..590P} to estimate the orbital eccentricity of the system, $>$ 61 degrees (Figure~\ref{fig7}).

From the H-R diagram we estimate an age of 3.3 $^{+2.2}_{-1.3}$ Myr. All three components of \ntts\ fall very nearly on the same isochrone, well within 1$\sigma$ of each other. Without adjusting for the multiplicity of the system, \citet{1999AJ....117.2381P} estimated a system luminosity of log(L/L$_\odot$) = -0.398. Using the initial visible light spectral type of M3 from W94, and the corresponding temperature T$_{eff}$ = 3451~K, \citet{1999AJ....117.2381P} derived an age estimate of 0.5 Myr. Taking the effects of multiplicity into account, they derived a luminosity error that increased their age estimate to 1-3 Myr, consistent with the age range that we have determined. The upper limit of our age estimate is also consistent with the 5-6 Myr age determined from the B star main sequence turnoff in Upper Scorpius \citep{1989A&A...216...44D}. 

\begin{figure*}
\includegraphics[angle=0,width=6.0in]{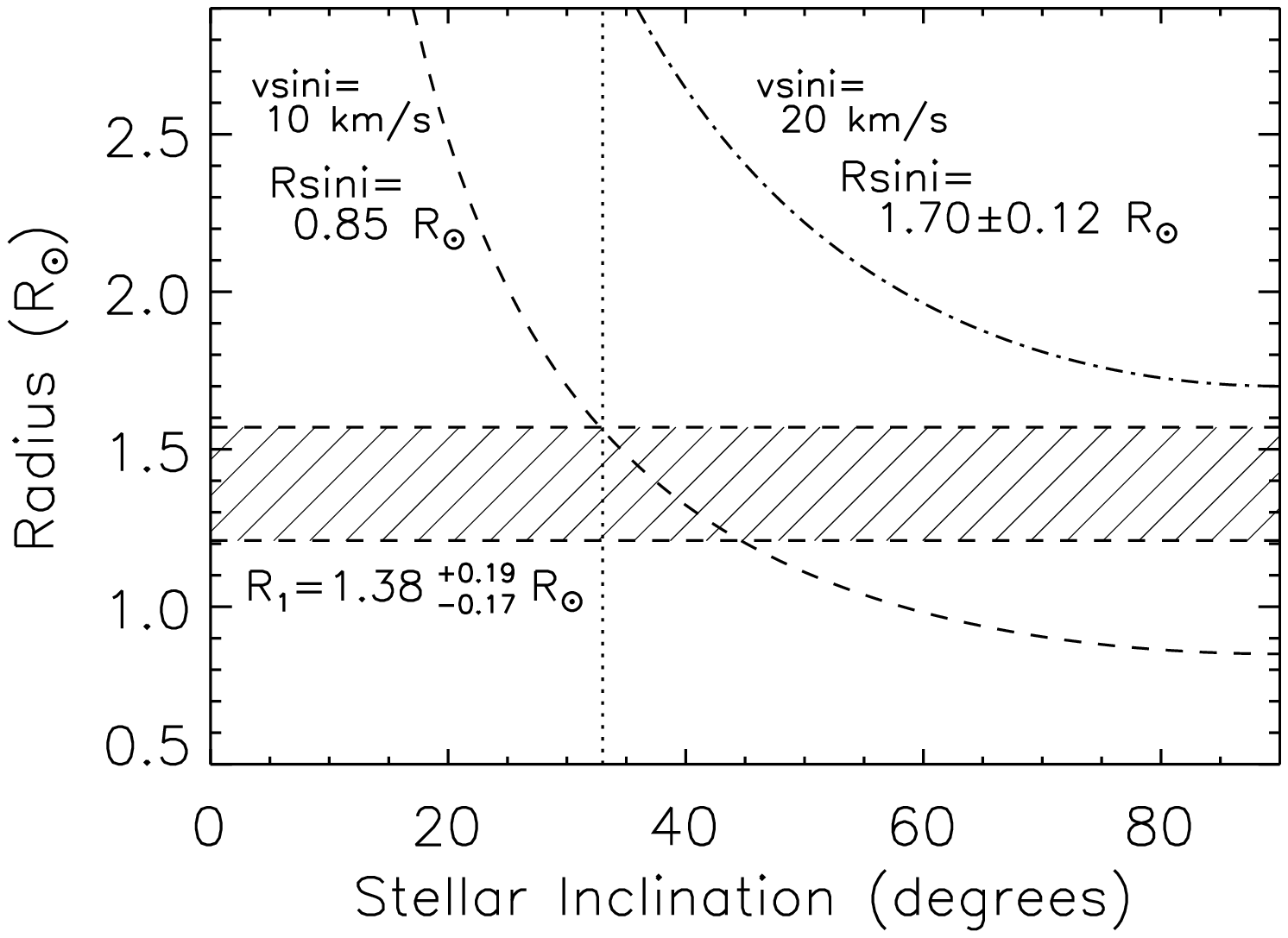}
\caption{Radius of the \ntts\ primary as a function of stellar inclination. The primary star's radius, derived from the luminosity and temperature used in Figure~\ref{fig6}, is over-plotted with 1$\sigma$ uncertainties (cross-hatched region). We determine Rsin$i$ from the rotational period reported by \citet{1998AJ....116..237A}, 4.3 $\pm$ 0.3 days, and our measured vsin$i$'s. Our IR spectra show the highest correlation with templates rotated to vsin$i$ $\approx$ 20 \kms; our visible light data yield vsin$i$ $\le$ 10~\kms. Our radius derived for the primary is consistent with values of Rsin$i$ for a range of $v$sin$i$ of 10$-$20 km/s if the stellar inclination is $>$33 degrees.} 
\label{fig8}
\end{figure*}

We use the primary star luminosity and T$_{eff}$ to derive the radius, R$_1$=1.38$^{+0.19}_{-0.17}$ R$_\odot$, which is shown in Figure~\ref{fig8}. Curves of Rsin$i$ are overplotted for values of $v$sin$i$ of 10 and 20 km/s.  The resulting Rsin$i$ curve for $v$sin$i=20$ km/s does not intersect with the derived radius of 1.38 R$_{\odot}$, suggesting that the lower value for vsin$i$ is more accurate.

\subsection{Possibility of SB Eclipse}
To estimate the probability of at least a partial eclipse, we assume that the primary and secondary stars have equal radii, 1.38$^{+0.19}_{-0.17}$ R$_\odot$, and use the value determined in our orbital fit for the semi-major axis, asin$i$ = 25.12 $\pm$ 0.25 R$_\odot$ (Table~\ref{tbl-4}). For the plausible range of stellar radii we expect a partial eclipse for inclinations $>$ 83$^o$, thus for an orbital inclination of $>$61 degrees (Figure 7) we derive an eclipse probability of $\sim$24\%.

Dedicated photometric monitoring of \ntts\ could reveal eclipses, and thus a value for the orbital inclination, providing absolute masses. The absence of eclipses would also be useful in limiting the orbital inclination of the system and thus our mass estimates. Brightness variations arising from the spotted primary star's rotation, $\Delta$V = 0.164 magnitudes \citep{1998AJ....116..237A}, might complicate photometric observations of partial eclipses, although spot- and eclipse-induced variability have distinct signatures. Another difficulty in measuring an eclipse is the nearly integer period of the SB orbit, $\sim$17 days, resulting in limited phase coverage for a single observing season.

\subsection{Tertiary Orbital Information}

For the tertiary we have two astrometric measurements from the literature, \citep{2000A&A...356..541K,2001A&A...369..249W}, and another pair determined from our NIRSPEC+AO observations. Although the saved SCAM images taken on our AO nights were overexposed and saturated, the finite width of the slit and the separated SB and tertiary continua allow us to estimate the position angle\footnote{See: http://www2.keck.hawaii.edu/inst/tools/skypa/} and separation for two epochs. Table~\ref{tbl-6} lists the astrometric data for the tertiary, which show no significant change in the separation and only a small change in the position angle over the nearly 9 years of observation. Assuming a distance to Sco-Cen of 145 pc, the average on sky separation of 0$\farcs$21 results in a projected separation of $\sim$30~AU. For this separation, the masses derived in \S 5.1, and assuming an edge-on, circular orbit, the space velocity of the tertiary relative to the SB center-of-mass would be $\sim$5.5 \kms\ and the minimum orbital period would be $\sim$143 years. Averaging our IR tertiary velocities and computing the velocity separation between the tertiary and the SB center-of-mass produces $\Delta$RV = -5.73 $\pm$ 0.44 \kms\ at the median epoch of our IR data. This velocity separation suggests that the tertiary orbit is either circular with a high inclination (implying near coplanarity with the SB) or eccentric with any possible inclination; \citet{2007prpl.conf..395M} show that systems with higher multiplicity are more likely to be nonaligned. Although we determined d$\gamma$/dt = -0.25 $\pm$ 0.04 (\kms\thinspace yr$^{-1}$) from the SB orbital solution, no significant trend is found in the tertiary radial velocities (d$v_T$/dt = 0.07 $\pm$ 0.09 \kms\thinspace yr$^{-1}$). One explanation for a missing trend might be that the tertiary is itself a binary, and thus more massive than we expect from a M2.5 alone. Yet, we see no evidence in the tertiary spectrum for another star. An alternative explanation for the d$\gamma$/dt that we have measured is a mismatch of the instrumental zero points. However, this is unlikely since both the visible and IR spectra were reduced using synthetic spectral templates and the Earth's atmosphere as the radial velocity standard frame of reference. Besides monitoring the SB for eclipses, other future work will include occasional astrometric imaging and NIRSPEC+AO spectroscopy to separate the SB and tertiary and monitor their radial velocities. 

\section{Summary}

\ntts\ (ScoPMS 20) is a WTTS spectroscopic binary (SB) with a tertiary component at 0$\farcs$21. By combining radial velocities derived from visible light and infrared spectra, acquired between 1987 June and 2007 April, we have determined the orbital parameters of the SB. Our orbital solution results in a 16.9243 $\pm$ 0.0002 day period, center-of-mass acceleration d$\gamma$/dt = -0.25 $\pm$ 0.04 (\kms\thinspace yr$^{-1}$), eccentricity e = 0.113 $\pm$ 0.008, and mass ratio q = 0.78 $\pm$ 0.01. 

Through the use of adaptive optics in the near-IR we were able to disentangle the SB and tertiary spectra.
Synthetic templates produced consistent and precise radial velocities, both in visible light and in the IR.
Observed IR templates \citep[e.g., ][]{pra02b} provided estimates for the component spectral types
of M2, M3, and M2.5 for the primary, secondary, and tertiary, respectively.  Based on these spectral types,
effective temperatures of 3560, 3415, and 3488 K were estimated from \citet{2003ApJ...593.1093L}.

We found H-band flux ratios of $\alpha = 0.55 \pm 0.10$ and $\beta = 0.81 \pm 0.09$, which we used to convert near-IR magnitudes to component luminosities. Plotting the stars on the theoretical tracks of \citet{bar98} gave an age for the coeval system of 3.3$^{+2.2}_{-1.3}$ Myr. Comparison of the minimum mass from the orbital solution to the H-R diagram mass estimates limits the SB orbital inclination to $>$ 61$^o$. Combining T$_{eff}$ and L, we estimated the primary radius to be 1.38$^{+0.19}_{-0.17}$ R$_\odot$. Assuming a similar secondary radius, and taking into account the derived semi-major axis and limits on the SB orbital inclination, we predict a $\sim$24$\%$ chance of a partial eclipse of the SB. High cadence imaging of the system would reveal an SB eclipse which could help determine the inclination and thus yield component mass estimates.

\begin{acknowledgments}
We thank our collaborators for their contributions: T. Barman for producing our synthetic 
template library, and M. Simon for discussions on choosing 
the optimal cross-correlation templates. We sincerely thank the Keck 
staff for their support of this science, in particular T. Stickel, J. Rivera, S. Magee, G. Puniwai, 
C. Wilburn, G. Hill, and B. Schaefer. We are grateful to the anonymous referee
for a timely and succinct report. This research was funded in part by a NASA Keck PI Data Award, administered
by the NASA Exoplanet Science Institute, and NSF Grants AST-0444017 and AST-1009136 (to L.P.). 
GT acknowledges partial support from the NSF through grant AST-1007992. Some of the data described herein was taken
on the Keck II telescope with time granted by NOAO, through the Telescope
System Instrumentation Program (TSIP). TSIP is funded by NSF. 
This work made use of the SIMBAD reference database, the NASA
Astrophysics Data System, and the data products from the Two Micron All
Sky Survey, which is a joint project of the University of Massachusetts
and the Infrared Processing and Analysis Center/California Institute
of Technology, funded by the National Aeronautics and Space
Administration and the National Science Foundation. This publication makes use of data products 
from the Wide-field Infrared Survey Explorer, which is a joint project of the University of California, 
Los Angeles, and the Jet Propulsion Laboratory/California Institute of Technology, funded by
 the National Aeronautics and Space Administration.
Data presented herein were obtained at the W.M. Keck
Observatory from telescope time allocated to the National Aeronautics
and Space Administration through the agency's scientific partnership
with the California Institute of Technology and the University of
California. The Observatory was made possible by the generous
financial support of the W.M. Keck Foundation.
We recognize and acknowledge the
significant cultural role that the summit of Mauna Kea
plays within the indigenous Hawaiian community and are
grateful for the opportunity to conduct observations
from this special mountain.
\end{acknowledgments}

\clearpage

\pagestyle{empty}

\begin{deluxetable}{lr}
\tablewidth{0pt}
\tablecaption{Observed Properties of \ntts\
\label{tbl-1}}
\tablehead{\colhead{Property} & \colhead{Value} }
\startdata
R.A.\tablenotemark{a} (J2000) & 16:01:05.2\\
Dec.\tablenotemark{a} (J2000) & -22:27:31.2\\
V\tablenotemark{b} (mag) & 13.74 $\pm$ 0.03\\
J\tablenotemark{a}  (mag) & 9.74 $\pm$ 0.03\\
H\tablenotemark{a}  (mag) & 9.01 $\pm$ 0.02\\
K\tablenotemark{a}  (mag) & 8.75 $\pm$ 0.02\\
L$'$\tablenotemark{c} (mag) & 8.46 $\pm$ 0.09\\
M\tablenotemark{c} (mag) & 7.85 $\pm$ 0.35\\
W1\tablenotemark{d} (mag) & 8.64 $\pm$ 0.02\\
W2\tablenotemark{d}  (mag) & 8.53 $\pm$ 0.02\\
W3\tablenotemark{d}  (mag) & 8.43 $\pm$ 0.03\\
W4\tablenotemark{d}  (mag) & 8.21 $\pm$ 0.28\\
H$\alpha$ EW\tablenotemark{b,e} $~$(\AA) & $-$3.32 $\pm$ 0.72\\
Li EW\tablenotemark{b} (\AA) & 0.55 $\pm$ 10$\%$\\
\enddata

\tablenotetext{a}{2MASS All-Sky Point Source Catalog}
\tablenotetext{b}{\citet{1994AJ....107..692W}}
\tablenotetext{c}{\citet{1997AJ....114..301J}}
\tablenotetext{d}{WISE All-Sky Data Release}
\tablenotetext{e}{\citet{1994ARA&A..32..465M}}

\end{deluxetable}

\clearpage

\pagestyle{empty}

\begin{deluxetable}{lccccccl}
\tablewidth{0pt}
\tablecaption{Visible Light Spectroscopic Observations and Radial Velocities
\label{tbl-2}}
\tablehead{
\colhead{UT Date} & \colhead{HJD} &
\colhead{$v_1$}  & \colhead{$\sigma_{v1}$}  &\colhead{$v_2$ }  & \colhead{$\sigma_{v2}$}  &\colhead{$v_{tert}$ }  &\colhead{SB} \\
\colhead{$~$} & \colhead{(2,400,000$+$)} &
\colhead{(km s$^{-1}$)} & \colhead{(km s$^{-1}$)} & \colhead{(km s$^{-1}$)} & \colhead{(km s$^{-1}$)} & \colhead{(km s$^{-1}$)} & \colhead{Phase}}
\startdata
 1987 Jun 09 & 46955.7364  & -41.07 & 1.75 & 31.63 & 4.57 & -4.71 & 0.6551\\
 1988 May 29 & 47310.7294 & -35.85 & 0.81 & 30.64 & 2.13 & -6.27 & 0.6305\\
 1988 Jul 03 & 47345.7140  & -33.98 & 1.24 & 32.58 & 3.23 & -5.53 & 0.6976\\
 
 1989 Jan 19 & 47546.0505  & -31.65 & 1.75 & 27.24 & 4.57 & -3.74 & 0.5348\\
 1989 Jan 23 & 47550.0308  & -27.58 & 1.94 & 18.37 & 5.08 & -4.11 & 0.7700\\
 1989 Apr 12 & 47628.9749  & -20.97 & 1.87 & 10.95 & 4.89 & -11.84 & 0.4345\\
 
 1989 Apr 16 & 47632.9116  & -36.85 & 1.40 & 34.40 & 3.66 & -6.95 & 0.6671\\
 1989 Apr 19 & 47635.9830 & -9.80 & 2.47 & $~$ & $~$ & $~$ & 0.8486\\
 1989 May 12 & 47658.9375  & 19.60 & 2.85 & -35.74 & 7.47 & -5.70 & 0.2049\\
 
 1989 May 15 & 47661.8636  & -12.68 & 1.81 & 4.99 & 4.72 & -0.68 & 0.3778\\
 1989 May 18 & 47664.8394  & -32.58 & 1.46 & $~$ & $~$ & $~$ & 0.5536\\
 1989 May 22 & 47668.8128  & -24.29 & 1.49 & 13.80 & 3.90 & -1.60 & 0.7884\\
 
 1989 May 23 & 47669.8250  & -13.61 & 1.17 & $~$ & $~$ & $~$ & 0.8482\\
 1989 May 24 & 47670.7618  & 2.43 & 1.28 & -11.98 & 3.34 & -4.65 & 0.9036\\
 1989 May 25 & 47671.7267  & 12.96 & 1.20 & -34.44 & 3.14 & -7.22 & 0.9606\\
 
 1989 May 26 & 47672.7810  & 25.38 & 1.94 & -46.11 & 5.08 & -5.68 & 0.0229\\
 1989 Jun 25 & 47702.7097  & -21.89 & 1.81 & 13.28 & 4.72 & -3.30 & 0.7913\\
 1989 Jun 26 & 47703.6753  & -14.55 & 2.64 & $~$ & $~$ & $~$ & 0.8483\\
 
 1989 Jul 12 & 47719.6712  & -24.27 & 1.87 & 24.79 & 4.89 & -4.34 & 0.7934\\
 1990 Feb 06 & 47929.0414  & 24.93 & 1.17 & -48.96 & 3.05 & -7.29 & 0.1644\\
 1990 Jun 03 & 48045.7813  & 28.00 & 1.08 & -46.02 & 2.82 & -1.28 & 0.0621\\

 1990 Jun 11 & 48053.7827  & -32.20 & 1.20 & 23.17 & 3.14 & -2.70 & 0.5349\\
 1990 Jun 12 & 48054.7620  & -33.88 & 1.22 & 37.06 & 3.19 & -9.89 & 0.5928\\
 1991 Mar 31 & 48346.9125  & -11.90 & 1.35 & 1.82 & 3.52 & -11.89 & 0.8550\\
 
 1991 Apr 01 & 48348.0054  & 4.70 & 1.32 & -11.27 & 3.46 & -1.14 & 0.9195\\
 1991 Apr 02 & 48348.9374  & 18.10 & 1.30 & -34.79 & 3.40 & -7.49 & 0.9746\\
 1991 Apr 25 & 48371.9536  & -4.57 & 2.85 & -4.59 & 7.47 & -4.57 & 0.3346\\
 
 1991 May 01 & 48377.9066  & -36.38 & 1.87 & 41.07 & 4.89 & -8.52 & 0.6863\\
 1991 May 03 & 48379.8833  & -20.94 & 1.49 & 14.65 & 3.90 & 2.12 & 0.8031\\
 1991 May 21 & 48397.8579  & -10.87 & 2.85 & -1.15 & 7.47 & -1.21 & 0.8651\\
 
 1991 May 23 & 48399.7728  & 15.13 & 1.43 & -40.74 & 3.74 & -2.29 & 0.9783\\
 1991 Jun 03 & 48410.8182  & -38.65 & 1.24 & 34.68 & 3.23 & -5.73 & 0.6309\\
 1992 May 12 & 48754.8943  & 12.72 & 1.56 & -31.59 & 4.09 & -6.36 & 0.9612\\
 
 1992 Jun 09 & 48782.7909  & -36.16 & 1.24 & 30.45 & 3.23 & -6.97 & 0.6095\\
 1992 Sep 06 & 48871.6234  & -14.20 & 1.65 & 1.00 & 4.31 & -5.29 & 0.8583\\
 1993 Mar 31 & 49077.9395  & 29.73 & 1.65 & -54.00 & 4.31 & -8.79 & 0.0488\\
 1993 Apr 03 & 49080.9905  & 14.98 & 1.65 & -35.60 & 4.31 & -5.52 & 0.2290\\
\enddata

\end{deluxetable}

\clearpage

\pagestyle{empty}

\begin{deluxetable}{lccccccl}
\tablewidth{0pt}
\tablecaption{Infrared Spectroscopic Observations and Radial Velocities
\label{tbl-3}}
\tablehead{
\colhead{UT Date} & \colhead{HJD} &
\colhead{$v_1$}  & \colhead{$\sigma_{v1}$}  &\colhead{$v_2$ }  & \colhead{$\sigma_{v2}$}  &\colhead{$v_{tert}$ }  &\colhead{SB} \\
\colhead{$~$} & \colhead{(2,400,000$+$)} &
\colhead{(km s$^{-1}$)} & \colhead{(km s$^{-1}$)} & \colhead{(km s$^{-1}$)} & \colhead{(km s$^{-1}$)} & \colhead{(km s$^{-1}$)} & \colhead{Phase}}
\startdata
 2000 Jun 11\tablenotemark{a} & 51706.8238  & -12.69 & 2.54 & -4.26 & 2.67 & -4.26 & 0.3805\\
 2002 Jul 18 & 52473.7874  & -38.14 & 1.19 & 25.86 & 1.36 & -4.26 & 0.6977\\

 2003 Feb 08 & 52679.0828  & -20.60 & 2.63 & 4.90 & 3.00 & $~$ & 0.8279\\
 2003 Apr 20\tablenotemark{a} & 52750.1016  & 21.33 & 1.79 & -51.03 & 1.78 & -4.27 & 0.0242\\
 
 2004 Jan 27 & 53032.1879  & -38.17 & 1.64 & 25.83 & 3.28 & -5.17 & 0.6917\\
 2004 May 24\tablenotemark{b} & 53149.8809  & -38.32 & 1.69 & 29.78 & 1.61 & $~$ & 0.6457\\
 2007 Apr 30 & 54220.9531  & 3.17 & 2.10 & -28.82 & 4.21 & $~$ & 0.9317\\
\enddata

\tablenotetext{a}{NIRSPEC+AO Observation; SB and tertiary spectra separated.}
\tablenotetext{b}{NIRSPEC+AO Observation; SB only.}

\end{deluxetable}

\clearpage

\pagestyle{empty}

\begin{deluxetable}{lcccc}
\tablewidth{0pt}
\tablecaption{SB Orbital Elements and Derived Properties  
\label{tbl-4}}
\tablehead{\colhead{$~$} &  \colhead{Single-Lined} &  \multicolumn{3}{c}{Double-Lined} \\
\colhead{Element/Property} &  \multicolumn{1}{|c|}{Visible Light\tablenotemark{a}} & \colhead{Visible Light}  & \colhead{Infrared} & \multicolumn{1}{c|}{Combined}}
\startdata
P (days)&16.925 & $16.9257 \pm 0.0010$ & $16.9231 \pm 0.0023$ & $16.9243 \pm 0.0002$\\
$\gamma$ (\kms\ ) & -5.0 & -6.28 $\pm\ 0.25$ & -9.99 $ \pm\ 0.38$ & -8.06 $\pm\ 0.29$\\
d$\gamma$/dt (\kms\thinspace yr$^{-1}$) & $-$ & -0.01 $\pm\ 0.18$ & -0.04 $\pm\ 0.22$ & -0.25 $\pm\ 0.04$ \\
$K_1$ (\kms\ ) & $-$ & $33.06 \pm 0.38 $ & $32.39 \pm 2.01 $ & $33.12 \pm 0.33 $\\
$K_2$ (\kms\ ) & $-$ & $42.70 \pm 0.89 $ & $42.74 \pm 2.55 $ & $42.43 \pm 0.67 $\\
e & 0.10 & $0.115 \pm 0.010$ & $0.120 \pm 0.018$ & $0.113 \pm 0.008$\\
$\omega$ (degrees) & $-$ & $317.0 \pm 5.8$ & $322.1 \pm 33.6$ & $319.2 \pm 5.0$\\
T (MJD)&  $-$  & $ 52665.36\pm 0.33$ &  $52834.46 \pm 1.60$  & $52698.92 \pm 0.23$ \\
$M_1$ sin$^3 i$ & $-$ & $0.421 \pm 0.022$ & $0.414 \pm 0.076$ &$0.417 \pm 0.017$ \\
$M_2$ sin$^3 i$ & $-$ & $0.326 \pm 0.014$ & $0.314 \pm 0.059$ &$0.325 \pm 0.011$\\
$f$(m)\tablenotemark{b} & 0.051 & $-$ & $-$ & $-$ \\
$q = M_2/M_1$ & $0.71 \pm 0.05 \tablenotemark{c}$ & $0.77 \pm 0.02 $ & $0.76 \pm 0.07 $ & $0.78 \pm 0.01 $\\
$a_1$ sin $i$ (Gm) & 7.18 & $7.644 \pm 0.088$ & $7.482 \pm 0.463$ & $7.659 \pm 0.077$\\
$a_2$ sin $i$ (Gm) & $-$ & $9.872 \pm 0.207$  & $9.875 \pm 0.589$& $9.811 \pm 0.154$\\
$\chi_{reduced}^2$ & $-$ & 0.969 & 0.367 & 0.891\\
$N_1$ , $N_2$&  27\tablenotemark{c} , $-$ & 37 , 33 & 7 , 7 & 44 , 40\\
 $\overline\sigma_{v1}$ , $\overline\sigma_{v2}$  (\kms\ )& $-$ & 1.65 , 4.21 & 1.94 , 2.56 & 1.69 , 3.92\\ 
\enddata

\tablenotetext{a}{\citet{1994ARA&A..32..465M}}
\tablenotetext{b}{Mass function, $f$(m) = (m$_2$sin$i$)$^3$ / (m$_1$+m$_2$)$^2$}
\tablenotetext{c}{\citet{1994AJ....107..692W} identifies the three components and estimates the mass ratio from 27 visible light observations.}

\end{deluxetable}

\clearpage

\pagestyle{empty}

\begin{deluxetable}{lr}
\tablewidth{0pt}
\tablecaption{H-R Diagram Derived Properties of \ntts\
\label{tbl-5}}
\tablehead{\colhead{Property} & \colhead{Value} }
\startdata
SpT & M2/M3/M2.5\\
A$_\nu$(mag) & 0.5 $\pm$ 0.2\\
log(L$_1$/L$_\odot$) & -0.56 $\pm$ 0.04 \\
log(L$_2$/L$_\odot$) & -0.84 $\pm$ 0.04 \\
log(L$_3$/L$_\odot$) & -0.66 $\pm$ 0.04 \\
M$_1$ (M$_\odot$) & 0.50 $\pm$ 0.10 \\
M$_2$ (M$_\odot$) & 0.38 $\pm$ 0.10 \\
M$_3$ (M$_\odot$) & 0.44 $\pm$ 0.10 \\
M$_2$/M$_1$ & 0.76 $\pm$ 0.14\\
M$_3$/M$_1$ & 0.88 $\pm$ 0.14\\
M$_{T}$/M$_{SB}$ & 0.50 $\pm$ 0.17\\
R$_1$ (R$_\odot$) & 1.38$^{+0.19}_{-0.17}$\\
Age (Myr) & 3.3 $^{+2.2}_{-1.3}$\\
\enddata

\end{deluxetable}

\clearpage

\begin{deluxetable}{lcccc}
\tablewidth{0pt}
\tablecaption{Tertiary Astrometric Measurements
\label{tbl-6}}
\tablehead{\colhead{Date} & \colhead{Position Angle} & \colhead{$\sigma$$_{PA}$} & \colhead{Separation} & \colhead{$\sigma$$_{sep}$}\\
\colhead{(UT)} & \colhead{(degrees)} & \colhead{(degrees)} & \colhead{(\arcsec)} & \colhead{(\arcsec)}}
\startdata
1994.332\tablenotemark{a} & 313.7 &1.2 & 0.193 & 0.005\\
1998.348\tablenotemark{b} & 317.5 & 0.8 & 0.21 & 0.04\\
2000.444 & 315 & 4 & 0.22 & 0.04\\
2003.300 & 321 & 4 & 0.23 & 0.04\\
\enddata

\tablenotetext{a}{\citet{2000A&A...356..541K}}
\tablenotetext{b}{\citet{2001A&A...369..249W}}

\end{deluxetable}

\clearpage

\clearpage

\clearpage

\clearpage

\clearpage

\clearpage

\clearpage

\clearpage


\clearpage
\begin{thebibliography}{} 

\bibitem[Adams et al.(1998)]{1998AJ....116..237A} Adams, N.~R., Walter, 
F.~M., \& Wolk, S.~J.\ 1998, \aj, 116, 237

\bibitem[Baraffe et al.(1998)]{bar98} Baraffe, I., Chabrier, G.,
Allard, F., \& Hauschildt, P. H. 1998, \aap, 337, 403

\bibitem[Beck et al.(2003)]{2003ApJ...583..358B} Beck, T.~L., Simon, M., 
\& Close, L.~M.\ 2003, \apj, 583, 358 

\bibitem[Boden et al.(2005)]{2005ApJ...635..442B} Boden, A.~F., et al.\ 
2005, \apj, 635, 442

\bibitem[Browning et al.(2010)]{2010AJ....139..504B} Browning, M.~K., 
Basri, G., Marcy, G.~W., West, A.~A., \& Zhang, J.\ 2010, \aj, 139, 504 

\bibitem[Dahm et al.(2012)]{2012ApJ...745...56D} Dahm, S.~E.,
Slesnick, C.~L., \& White, R.~J.\ 2012, \apj, 745, 56

\bibitem[de Geus et al.(1989)]{1989A&A...216...44D} de Geus, E.~J., de Zeeuw, P.~T., \& Lub, J.\ 1989, \aap, 216, 44

\bibitem[Hauschildt et al.(1999)]{1999ApJ...512..377H} Hauschildt, P.~H., 
Allard, F., \& Baron, E.\ 1999, \apj, 512, 377

\bibitem[Jensen 
\& Mathieu(1997)]{1997AJ....114..301J} Jensen, E.~L.~N., \& Mathieu, R.~D.\ 1997, \aj, 114, 301

\bibitem[Jones et al.(2005)]{2005MNRAS.358..105J} Jones, H.~R.~A., 
Pavlenko, Y., Viti, S., et al.\ 2005, \mnras, 358, 105

\bibitem[K{\"o}hler et 
al.(2000)]{2000A&A...356..541K} K{\"o}hler, R., Kunkel, M., Leinert, C., \& Zinnecker, H.\ 2000, \aap, 356, 541

\bibitem[Kraus et al.(2008)]{2008ApJ...679..762K} Kraus, A.~L., Ireland, 
M.~J., Martinache, F., \& Lloyd, J.~P.\ 2008, \apj, 679, 762

\bibitem[Kurtz et al.(1992)]{1992ASPC...25..432K} Kurtz, M.~J., Mink, 
D.~J., Wyatt, W.~F., Fabricant, D.~G., Torres, G., Kriss, G.~A., 
\& Tonry, J.~L.\ 1992, Astronomical Data Analysis Software and Systems I, 25, 432

\bibitem[Lada(2006)]{2006ApJ...640L..63L} Lada, C.~J.\ 2006, \apjl, 640, 
L63 

\bibitem[Latham(1992)]{Latham:92}
 Latham, D.\ W. 1992, in IAU Coll.\ 135, Complementary Approaches to
Double and Multiple Star Research, ASP Conf.\ Ser.\ 32, eds.\ H.\ A.\
McAlister \& W.\ I.\ Hartkopf (San Francisco: ASP), 110

\bibitem[Latham et al.(2002)]{Latham:02}
 Latham, D.\ W., Stefanik, R.\ P., Torres, G., Davis, R.\ J., Mazeh,
 T., Carney, B.\ W., Laird, J.\ B., \& Morse, J.\ A. 2002, \aj, 124,
 1144

\bibitem[Luhman et al.(2003)]{2003ApJ...593.1093L} Luhman, K.~L., Stauffer, 
J.~R., Muench, A.~A., et al.\ 2003, \apj, 593, 1093

\bibitem[Mathieu(1994)]{1994ARA&A..32..465M} Mathieu, R.~D.\ 1994, \araa, 32, 465

\bibitem[Mathieu et al.(2000)]{2000prpl.conf..703M} Mathieu, R.~D., Ghez, 
A.~M., Jensen, E.~L.~N., \& Simon, M.\ 2000, Protostars and Planets IV, ed. V. Mannings, A.P. Boss \& S.S. Russell (Tucson: Univ. Arizona Press), 703

\bibitem[Mazeh et al.(2002)]{maz02} Mazeh, T., Prato, L., Simon, M.,
Goldberg, E., Norman, D., \& Zucker, S. 2002, \apj, 564, 1007

\bibitem[Mazeh et al.(2003)]{2003ApJ...599.1344M} Mazeh, T., Simon, M., 
Prato, L., Markus, B., \& Zucker, S.\ 2003, \apj, 599, 1344 

\bibitem[McLean et al.(1998)]{mcl98} McLean, I. S., et al. 1998,
SPIE, 3354, 566

\bibitem[McLean et al.(2000)]{mcl00} McLean, I. S., Graham, J. R.,
Becklin, E. E., Figer, D. F., Larkin, J. E., Levenson, N. A., \& Teplitz,
H. I. 2000, SPIE, 4008, 1048

\bibitem[Monin et al.(2007)]{2007prpl.conf..395M} Monin, J.-L., Clarke, 
C.~J., Prato, L., \& McCabe, C.\ 2007, Protostars and Planets V, 395

\bibitem[Palla \& Stahler(2001)]{2001ApJ...553..299P} Palla, F., \& Stahler, S.~W.\ 2001, \apj, 553, 299

\bibitem[Prato(1998)]{1998PhDT........16P} Prato, L.~A.\ 1998, 
Ph.D.~Thesis, SUNY Stony Brook

\bibitem[Prato et al.(2001)]{2001ApJ...549..590P} Prato, L., et al.\ 2001, 
\apj, 549, 590

\bibitem[Prato et al.(2002a)]{pra02a} Prato, L., Simon, M., Mazeh, T., Zucker, S., \&  McLean, I. S. 2002a, \apjl, 579, L99

\bibitem[Prato et al.(2002b)]{pra02b} Prato, L., Simon, M.,
Mazeh, T., McLean, I. S., Norman, D., \& Zucker, S. 2002b, \apj, 569, 863

\bibitem[Prato et al.(2003)]{2003ApJ...584..853P} Prato, L., Greene, T.~P., 
\& Simon, M.\ 2003, \apj, 584, 853

\bibitem[Prato(2007)]{2007ApJ...657..338P} Prato, L.\ 2007, \apj, 657, 338

\bibitem[Preibisch 
\& Zinnecker(1999)]{1999AJ....117.2381P} Preibisch, T., \& Zinnecker, H.\ 1999, \aj, 117, 2381

\bibitem[Press et al.(1992)]{1992nrfa.book.....P} Press, W.~H., Teukolsky, 
S.~A., Vetterling, W.~T., \& Flannery, B.~P.\ 1992, Numerical Recipes in Fortran: The Art of Scientific Computing, (2nd edn.; Cambridge: Cambridge Univ. Press)

\bibitem[Simon(1997)]{1997ApJ...482L..81S} Simon, M.\ 1997, \apjl, 482, L81

\bibitem[Steffen et al.(2001)]{2001AJ....122..997S} Steffen, A.~T., et al.\ 
2001, \aj, 122, 997 

\bibitem[Torres et al.(2007)]{Torres:07}
 Torres, G., Latham, D.\ W., \& Stefanik, R.\ P. 2007, \apj, 662, 602

\bibitem[Walter et al.(1994)]{1994AJ....107..692W} Walter, F.~M., Vrba, 
F.~J., Mathieu, R.~D., Brown, A., \& Myers, P.~C.\ 1994, \aj, 107, 692

\bibitem[Wilson(1941)]{wil41} Wilson, O. C. 1941, \apj, 93, 29

\bibitem[Woitas et  al.(2001)]{2001A&A...369..249W} Woitas, J., K{\"o}hler, R., \& Leinert, C.\ 2001, \aap, 369, 249

\bibitem[Woitas et al.(2001)]{2001A&A...376..982W} Woitas, J., Leinert, C., K{\"o}hler, R.\ 2001, \aap, 376, 982

\bibitem[de Zeeuw et al.(1999)]{1999AJ....117..354D} de Zeeuw, P.~T., 
Hoogerwerf, R., de Bruijne, J.~H.~J., Brown, A.~G.~A., 
\& Blaauw, A.\ 1999, \aj, 117, 354

\bibitem[Zucker \& Mazeh(1994)]{Zucker:94} Zucker, S., \& Mazeh,
T. 1994, \apj, 420, 806

\bibitem[Zucker et al.(1995)]{Zucker:95} Zucker, S., Torres, G., \&
Mazeh, T. 1995, \apj, 452, 863

\end{thebibliography}
\end{document}